\documentclass[conference]{IEEEtran}
\IEEEoverridecommandlockouts

\usepackage{cite}
\usepackage{listings}
\usepackage{array}
\usepackage{balance}
\usepackage{multirow}
\usepackage{pbox}
\usepackage{url}
\usepackage{paralist}
\usepackage{xspace}
\usepackage{booktabs}
\usepackage{amsmath,amssymb,amsfonts}
\usepackage{algorithmic}
\usepackage{graphicx}
\usepackage{subfigure}
\usepackage{textcomp}
\usepackage{todonotes}
\usepackage{xcolor}
\usepackage{balance}
\usepackage{booktabs}
\usepackage{orcidlink}

\newcommand\orcid[1]{\orcidlink{#1}~\text{#1}}
\makeatletter
\newcommand{\linebreakand}{%
  \end{@IEEEauthorhalign}
  \hfill\mbox{}\par
  \mbox{}\hfill\begin{@IEEEauthorhalign}
}
\makeatother

\newcommand{\pp}[1]{\vspace{6pt}\noindent\textbf{\emph{#1 --}}\xspace}

\def\BibTeX{{\rm B\kern-.05em{\sc i\kern-.025em b}\kern-.08em
    T\kern-.1667em\lower.7ex\hbox{E}\kern-.125emX}}

\begin{document}

\title{
    Pseudonymization at Scale:\\OLCF's Summit Usage Data Case Study
    \thanks{This manuscript has been authored by UT-Battelle, LLC, under contract DE-AC05-00OR22725 with the US Department of Energy (DOE). The publisher acknowledges the US government license to provide public access under the DOE Public Access Plan (http://energy.gov/downloads/doe-public-access-plan).}
}

\author{
    \IEEEauthorblockN{Ketan Maheshwari}
    \IEEEauthorblockA{\textit{NCCS/OLCF} \\
    \textit{Oak Ridge National Laboratory}\\
    Oak Ridge, TN, USA \\
    \orcid{0000-0001-6852-5653}}
    \and
    \IEEEauthorblockN{Sean R. Wilkinson}
    \IEEEauthorblockA{\textit{NCCS/OLCF} \\
    \textit{Oak Ridge National Laboratory}\\
    Oak Ridge, TN, USA \\
    \orcid{0000-0002-1443-7479}}
    \and
    \IEEEauthorblockN{Alex May}
    \IEEEauthorblockA{\textit{NCCS/OLCF} \\
    \textit{Oak Ridge National Laboratory}\\
    Oak Ridge, TN, USA \\
    \orcid{0000-0002-6258-5302}}
    \linebreakand
    \IEEEauthorblockN{Tyler Skluzacek}
    \IEEEauthorblockA{\textit{NCCS/OLCF} \\
    \textit{Oak Ridge National Laboratory}\\
    Oak Ridge, TN, USA \\
    \orcid{0000-0003-2242-4931}}
    \and
    \IEEEauthorblockN{Olga A. Kuchar}
    \IEEEauthorblockA{\textit{NCCS/OLCF} \\
    \textit{Oak Ridge National Laboratory}\\
    Oak Ridge, TN, USA \\
    kucharoa@ornl.gov}
    \and
    \IEEEauthorblockN{Rafael Ferreira da Silva}
    \IEEEauthorblockA{\textit{NCCS/OLCF} \\
    \textit{Oak Ridge National Laboratory}\\
    Oak Ridge, TN, USA \\
    \orcid{0000-0002-1720-0928}}
%
}

\maketitle

\begin{abstract}
The analysis of vast amounts of data and the processing of complex 
computational jobs have traditionally relied upon high performance computing 
(HPC) systems, which offer reliable and efficient management of large-scale
computational and data resources. Understanding these analyses' needs is
paramount for designing solutions that can lead to better science, and
similarly, understanding the characteristics of the user behavior on those
systems is important for improving user experiences on HPC systems. A common
approach to gathering data about user behavior is to extract workload
characteristics from system log data available only to system administrators.
Recently at Oak Ridge Leadership Computing Facility (OLCF), however, we
unveiled user behavior about the Summit supercomputer by collecting data from a
user's point of view with ordinary Unix commands.

In this paper, we discuss the process, challenges, and lessons learned while
preparing this dataset for publication and submission to an open data
challenge. The original dataset contains personal identifiable information
(PII) about the users of OLCF which needed be masked prior to publication, and
we determined that anonymization, which scrubs PII completely, destroyed too
much of the structure of the data to be interesting for the data challenge. We
instead chose to pseudonymize the dataset, which reduced the linkability of
the dataset to the users' identities. Pseudonymization is significantly more
computationally expensive than anonymization, and the size of our dataset,
which is approximately 175 million lines of raw text, necessitated the
development of a parallelized workflow that could be reused on different HPC
machines. We demonstrate the scaling behavior of the workflow on two leadership
class HPC systems at OLCF, and we show that we were able to bring the overall
makespan time from an impractical 20+ hours on a single node down to around 2
hours. As a result of this work, we release the entire pseudonymized dataset
and make the workflows and source code publicly available.


\end{abstract}


\section{Introduction}
\label{sec:intro}
Understanding resource usage of high performance computing (HPC) systems is 
paramount for designing solutions that help perform better science, be it
through the efficiency of job and I/O throughput, energy consumption, or 
other considerations~\cite{mann2015allocation, hou2022prediction}.
Traditionally, workload trace archives have gathered and published datasets 
that capture key features of HPC resource usage~\cite{feitelson2014experience, 
javadi2013failure}. Currently, most of these available workload traces provide 
coarse-grained metrics (e.g., number of jobs submitted, requested number of 
cores, timestamp of submissions, etc.). These traces and their respective 
gathered metrics have been extensively used by researchers\footnote{To date,
200+ bibliographical references have been made to the Parallel Workload 
Archive~\cite{feitelson2014experience}.}. The method for using these traces 
typically involves a simulation which replays the jobs' arrival with determined 
computing requirements so that schedulers' efficiency can be assessed.
Some studies have used these traces to identify characteristics of user 
behavior using HPC systems~\cite{schlagkamp2016consecutive, 
paul2020understanding}. These studies have focused on discovering users' jobs 
submission patterns to improve load balancing and fairness. With the increasing 
usage of machine learning techniques for improving the performance of HPC
systems and their applications, however, there is a need for capturing 
fine-grained data that span multiple services at different levels of the HPC 
software stack.

More recently, we have collected and analyzed observational data on the login 
nodes from the Summit leadership class supercomputer hosted at the Oak Ridge 
Leadership Computing Facility (OLCF) at Oak Ridge National Laboratory 
(ORNL)~\cite{wilkinson2022iccs}. By periodically sampling user activities (job 
queues, running processes, etc.), we were able to unveil key usage patterns 
that evidence misuse of the system, including gaming the policies, impairing 
I/O performance, and using login nodes as a sole computing resource. However,
gathering and publishing such a dataset is a challenging undertaking. First,
data are collected using several tools, each with their own data formats and
granularity. Second, data gathered from two different tools may provide
overlapping (non-conformant) information. Third, in order to make this dataset 
openly available to the community, it is necessary to remove any Personal 
Identifiable Information (PII) from the dataset. In our previous 
work~\cite{wilkinson2022iccs}, we have focused on the first two of the
aforementioned challenges, but because the data were limited to processing and
analysis within our institution, no PII masking procedures were necessary
during these activities. These procedures did become necessary, however, when
we decided to publish the dataset and submit it as part of the Smoky Mountain
Conference Data Challenge \cite{smcdc2022-website}.

In this paper, we address the challenges of the scientific data management 
lifecycle for curating, redacting, and publishing a dataset about usage of
OLCF's Summit supercomputer to the research community. We chose to redact the
data using pseudonymization rather than anonymization for reasons we will
detail. In short, pseudonymization preserves more structure in the resulting
dataset, allowing a wider range of questions to be answered by careful analysis
of the dataset. Anonymization, for example, would still allow the calculation
of simple aggregate statistics like total batch jobs submitted or an average
number submitted per day; pseudonymization allows these calculations to be
binned by users and/or projects, without revealing identities. Insights from
the community could be seamlessly mapped by our system administrators to actual
users, software, and projects, for example, in order to improve policies,
enhance support, or even influence design of subsequent resources. Here, we
specifically present a data lifecycle management use case that describes the
end-to-end process from data gathering and mining to large-scale processing and
the open release of the dataset.

This Summit user behavior dataset~\cite{summitdata2022} comprises a collection
of samples recorded by a set of system tools that capture usage on the login
nodes with ordinary Unix commands, once every hour, from January 1, 2020 to
December 31, 2021. The published dataset is composed of more than 3,500 files
representing more than 175 million lines of raw text and accounting for more
than 20~GB of storage volume. Producing such a dataset entails the following
lifecycle: (i)~acquisition/generation or collection of data; (ii)~organization,
preprocessing, screening, and filtration; (iii)~analysis, analytics, and
processing of organized data; (iv)~publication of results obtained from
analysis; (v)~preparation and redaction of data for release through
anonymization and/or pseudonymization before subsequent packaging; and
(vi)~release and post-release management. In this work, we describe the
challenges faced at each stage of this lifecycle with emphases on the
pseudonymization process of the dataset and the operational process for
complying with the laboratory's institutional policies. Specifically, this work
makes the following contributions:

\begin{enumerate}
    \item We describe a set of observational data from the login nodes of the 
          leadership-class Summit supercomputer at OLCF;
    \item We present two open-source, reusable, and portable large-scale
          scientific workflows for anonymizing and pseudonymizing the dataset;
    \item We quantify the efficiency of each workflow in terms of scalability
          on two leadership-class supercomputers at OLCF;
    \item We discuss the implications of institutional policies that may
          severely impact the data management lifecycle, both in terms of
          dataset processing complexity and timelines.
\end{enumerate}

Note that in this paper we do not intend to draw conclusion from the dataset;
instead our goal is to describe the data lifecycle management for gathering,
processing, and releasing the dataset. We refer the reader to our previous 
work~\cite{wilkinson2022iccs} with this dataset, in which we have presented
findings regarding key usage patterns that we believe will shed light on the
usage of login nodes on contemporary clusters and supercomputers.

This paper is organized as follows. Section~\ref{sec:relatedwork} brings in the
context by discussing other similar recent efforts surrounding the topic of
data lifecycle management and release. Section~\ref{sec:dataset} describes the
dataset in detail to help the reader inspect and understand the various
elements in the dataset. Section~\ref{sec:workflow} discusses the need and
evolution of the anonymization and pseudonymization and dives deeper into its
nuances including scalability performance studies on two leadership class 
HPC systems. Section~\ref{sec:lessonslearned} discusses technical and 
non-technical challenges and lessons learned. Section~\ref{sec:conclusion} 
concludes with a summary of results and perspectives on future work.

\section{Related Work}
\label{sec:relatedwork}

The work presented in this paper surrounds two main areas of research. First,
the analysis, analytics, and release of large-scale datasets and, second, the
process of data pseudonymization and associated workflows that perform this
process. This section discusses relevant related research in both of these
areas.

The activity of pseudonymization is most prominently practiced in the medical
research and results dissemination in order to protect the Personal
Identifiable Information (PII) of participating individuals. The practice also
underlies the implications of regulations, in particular General Data
Protection Regulation (GDPR)~\cite{gdpr}. In fact, it could be argued that GDPR
compliant pseudonymization is the standard that curators and repositories need
to adhere to ensure the safe sharing of information now, and in the future.
Most recent examples include Covid-19 related datasets that have been made 
public available online~\cite{covidpseudo}. The community is also
cognizant of the associated software complexities in the face of pseudonymizing
datasets that are distributed~\cite{Kohlmayer2019}. In our case, we do think
the trouble of going through the pseudonymization is justified for the reasons
we discussed in section~\ref{sec:intro}.


The work presented in~\cite{mitdata} involves a complete \emph{anonymization} of
data. The authors have preferred to keep the process of anonymization as
confidential. We chose to publish the process as a reusable, portable, and
repurposable workflow for the community to take advantage of. To the best 
of our knowledge, this is the first work that provides an HPC user-focused 
pseudomized dataset, an open description of the process, and a reusable, 
portable, and repurposable workflow.

Other similar datasets that have been published are listed below:
\begin{enumerate} 
\item Blue Waters data~\cite{bluewaters} which is the result of scientific 
      data processing since 2012 on the Blue Waters supercomputer. The data is 
      fully anonymized. 
\item Google Clusters Data~\cite{clusterdata:Wilkes2020,clusterdata:Wilkes2020a}, 
      a small (7-hour) sample of resource-usage information from a Google 
      production cluster in 2010. This trace primarily provides data about 
      resource requests and usage. It contains no information about end users, 
      their data, or access patterns to services.
\item Parallel Workload Archive~\cite{feitelson2014experience}, a popular fully
      anonymized dataset of HPC schedulers from a collection of HPC facilities
      worldwide. The dataset is comprised of job scheduling data including 
      jobs requirements (number of cores, walltime, etc.) and user 
      identification.
\end{enumerate}

As datasets become too large for data curators to behold at once, further
research and work with pseudonymization could become part of a ``curators'
toolkit" that will help ensure that datasets meet increasingly stricter data
privacy regulations while. This is particularly important as it is becoming
acknowledged within the curatorial community that when datasets become too
large most curators are ``spot-checking" subsets of files which could become a
security risk~\cite{advpseu}. Our work could help mitigate these risky practices
and would be a welcome addition to curatorial workflows. 

\section{OLCF's Summit Usage Dataset}
\label{sec:dataset}

The dataset~\cite{wilkinson2022iccs, summitdata2022} is comprised of 
observational data collected on the login nodes from the Summit leadership 
class supercomputer hosted at the Oak Ridge Leadership Computing Facility 
(OLCF). The dataset contains Summit's login nodes' performance (CPU, memory,
and disk usage) and users activity (logged-in users, the programs they were
running on the login nodes, and the status of all user jobs). The data is
collected at every hour since January 1, 2020. To date, the dataset represents
activity from 1,967 unique users, who submitted 1,783,867 jobs, of which
1,073,754 completed successfully while 705,103 had a non-zero exit code.

The data is collected using a shell script running in a \texttt{while} loop
within a \texttt{tmux} session on each of the five Summit login nodes.
The data consists of one file per login node per day organized into directories 
following a ``MonYYYY" naming convention (e.g., Jun2020). Currently, the dataset
is composed of 3,500+ files that accounts for about 20~GB total data footprint.
Each file may range between 1--15~MBs. In total, the dataset has 175,236,847 
lines of text.
Although data collection is performed continuously, some data may be incomplete
due to Summit being under planned maintenance or unavailable due to external 
factors (e.g., network outages), or to glitches in the data collection process.
Login nodes usage performance and users data in each file is organized into 
sections and subsections. Each section comprises data gathered within an hour
of the day (i.e., there are 24 sections bound marked by the hour and ``endsnap" 
in each file---with exceptions where the process was abruptly interrupted).
Each hourly section consists of 10 subsections containing the following data:

\begin{enumerate}
    \item The output of the Unix \texttt{w} command; 
    \item The contents of \texttt{/proc/meminfo} file; 
    \item The contents of \texttt{/proc/vmstat} file; 
    \item The output of the Unix \texttt{ps aux} command (excluding root owned processes); 
    \item The output of the Unix \texttt{top} command (excluding root processes); 
    \item Information on all the jobs currently active in scheduler; 
    \item Time span to run unaliased \texttt{ls} command in \texttt{\$HOME};
    \item Time span to run a colored \texttt{ls} command in \texttt{\$HOME};
    \item Time span to create a 1~GB file in General Parallel File System (GPFS) scratch;
    \item Output of the Unix \texttt{df -h} command excluding the \texttt{tmpfs} filesystems.
\end{enumerate}

\medskip
Note that the output of the above commands and file contents may provide 
overlapping, non-conformant, or aggregated information. Therefore, conclusions 
drawn from the dataset should carefully account for these conditions.





\section{Anonymization and Pseudonymization of Personal Identifiable Information}
\label{sec:workflow}

Both anonymization and pseudonymization refer to protecting the confidentiality 
of personal identifiable information (PII)~\cite{stalla2016anonymous}. While 
the anonymization process attempts to protect the data in such a way that the 
personal data can no longer be identified, pseudonymizing entails in processing 
PII data in such a way that the data can no longer be attributed to a specific 
data subject without the use of additional information. In this work, we favor 
the latter as we seek to incorporate potential new insights unveiled through 
the analysis of the dataset by the community.

The drawback of the above choice is that the process for pseudonymizing a 
dataset is far more complex and computationally expensive than performing a 
plain anonymization. In this section, we explore both approaches to contrast
the computational complexity and costs associated with each alternative.

\subsection{Anonymization Workflow}

\begin{figure}[!t]
    \centering
    \includegraphics[width=\linewidth]{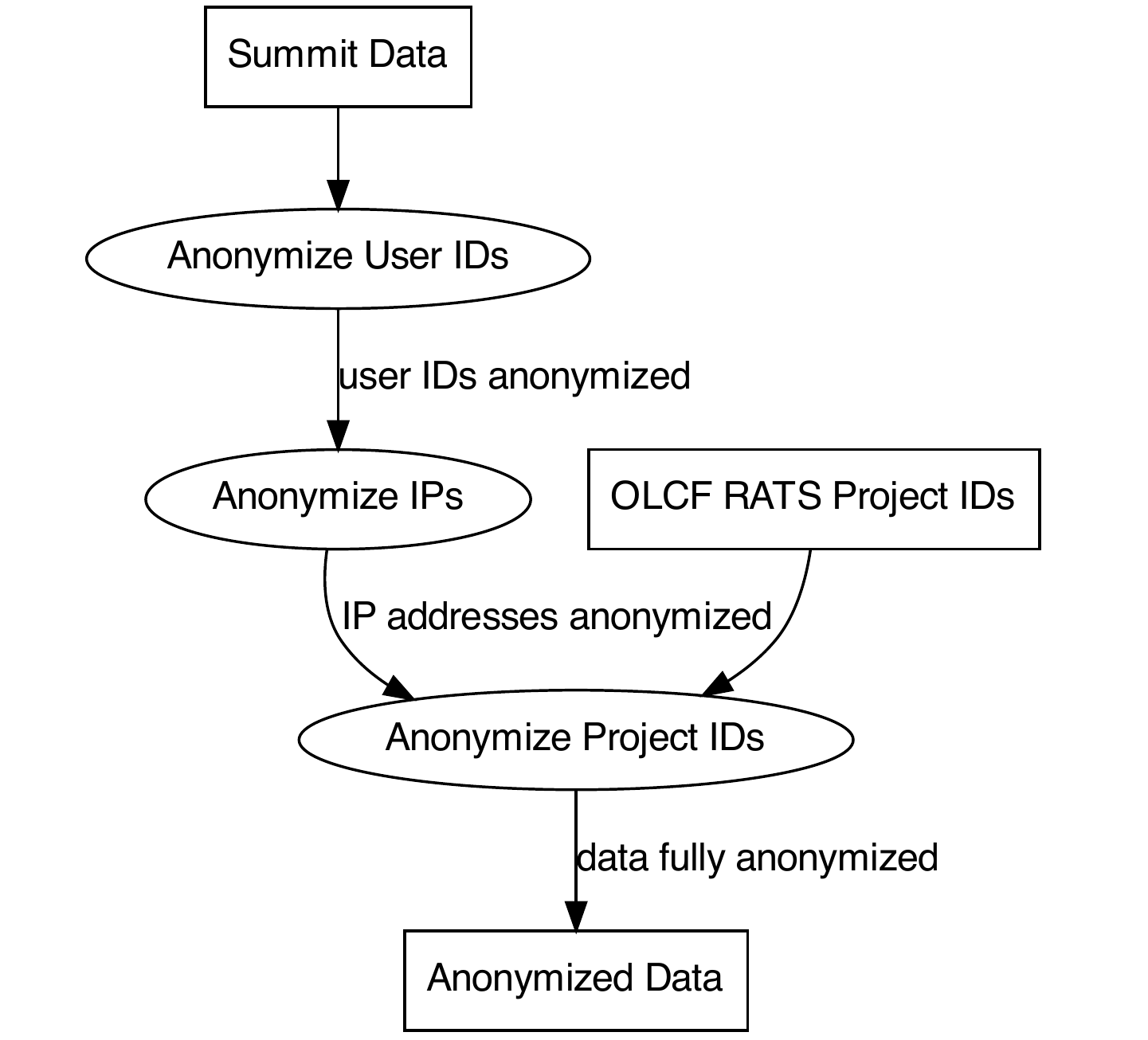}
    \caption{An abstract representation of the anonymization workflow. An instantiation of this workflow would produce series of pipelines (the three stages represented in oval shape) for each file to be analyzed.}
    \label{fig:anonworkflow}
\end{figure}

The anonymization process is described as a scientific reusable workflow 
(Figure~\ref{fig:anonworkflow}) in which the data collected from Summit's login
nodes (hereafter called ``Summit data" for short) follows a PII removal process
that consists on anonymizing user identifications, IP addresses, and project 
names/identities. The first stage in the workflow performs the extraction of 
user IDs from the data. This extraction is performed by running a Unix shell 
script with common utilities such as \texttt{awk} to identify known places 
where user IDs appear and replace them by random generated IDs (e.g., UUIDs).
In the second stage, the workflow seeks for IP addresses by replacing matching
regular expression patterns by random generated strings. In the last stage, the
workflow perform substitutions of project IDs. To this end, we feed the 
workflow with an external database that provides a list of OLCF's project IDs
(the RATS customer relationship management tool). Although this last step of
the workflow is relatively tied to an OLCF system, we claim that the proposed
workflow is generalizable; for example, RATS could be replaced by a simple 
seek-and-replace operation using regular expression patterns or any other 
customer relationship management tool.

We have implemented the anonymization workflow as a 
Swift/T~\cite{wozniak2013swift} workflow application. Swift/T workflows are 
compiled into MPI programs that are optimized for running at scale on HPC 
clusters. We run the workflow on the ORNL's Summit leadership class HPC 
system~\cite{vazhkudai2018design}. Summit is equipped with 4,608 compute nodes, 
in which each is equipped with two IBM POWER9 processors (42 cores), six NVIDIA 
Tesla V100 accelerators each with 96 GiB of HBM2, 512 GB of DDR4 memory, and 
connection to a 250 PB GPFS scratch filesystem. The workflow implementation 
and all analyses scripts are available on GitHub~\cite{Maheshwari_loginanalysis_2022}.

We ran an instance of the workflow to process the entire Summit data on 100 
Summit nodes. Each CPU core mostly processed a single file from the dataset 
and performed the necessary anonymization transformations for removing any PII.
(Note that different stages of the workflow may not necessarily run in the same
CPU or node for a single file, as the workflow scheduler may seek for available
slots as the next workflow task becomes ready, i.e. all its dependencies have 
been satisfied. This shuffling of tasks may yield added overhead as it may 
trigger data movement or additional I/O operations throughout the workflow 
execution.) For this experiment, the workflow makespan (i.e., overall execution 
time) is 1,661s, which is relatively low when considering the high number of 
files and substitutions to perform. This result emphasizes that the efficiency
of the anonymization workflow is mostly driven by the number of files to be 
processed (which is not the sole factor impacting the pseudonymization workflow
as seen below).

\subsection{Pseudonymization Workflow}

\begin{figure}[!t]
    \centering
    \includegraphics[width=\linewidth]{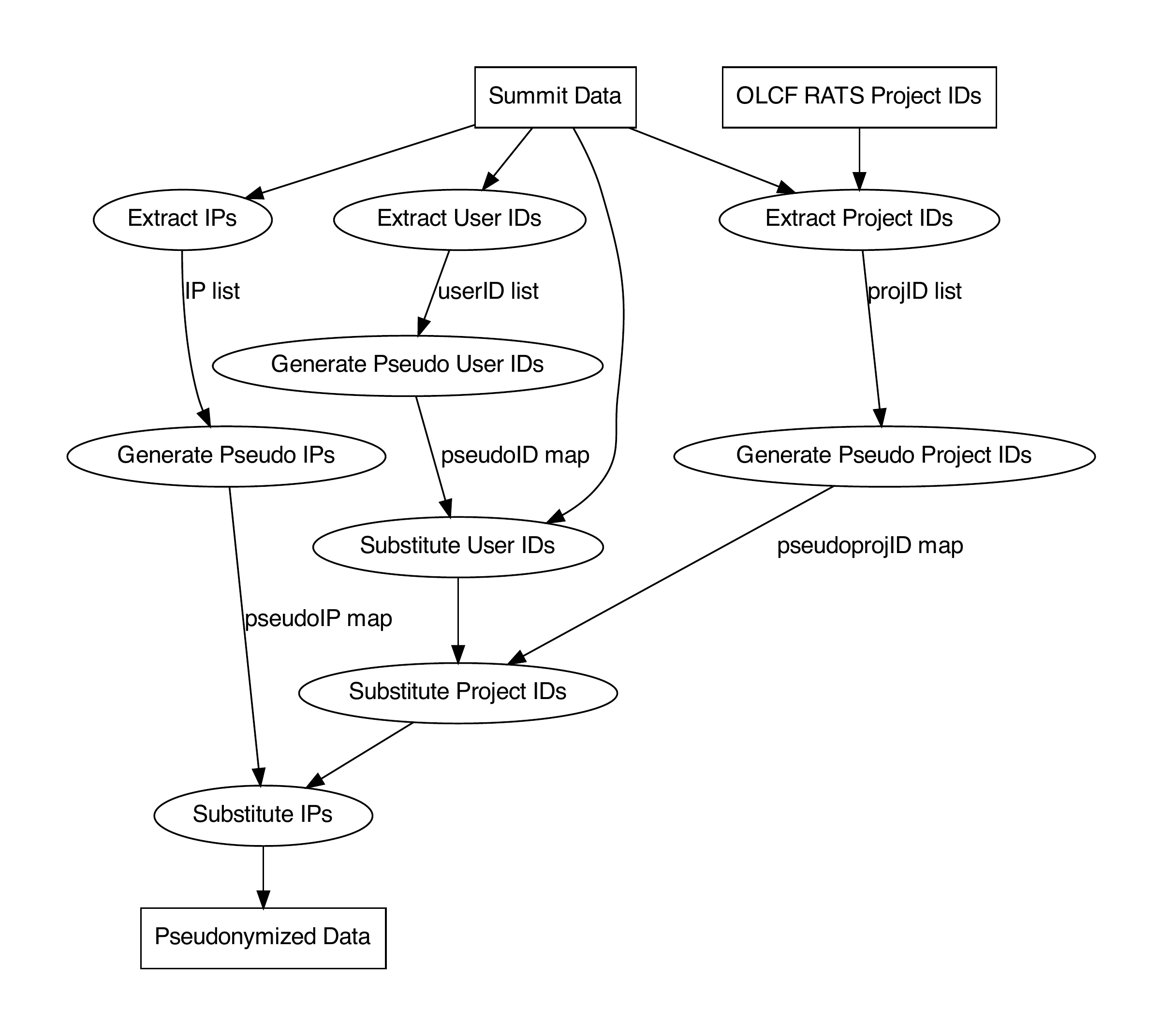}
    \caption{An abstract representation of the pseudonymization workflow. An instantiation of this workflow would produce series of branches (i.e., the three pipelines composed of stages represented in oval shape) for each file that would be analyzed.}
    \label{fig:workflow}
\end{figure}

The pseudonymization process is also implemented as a reusable scientific 
workflow, however its complexity is severely increased due to added stages that 
in addition to exchanging personal data with non-identifying data, it also 
needs to generate and maintain a ``map'' of information to recreate the 
original data. Similarly to the anonymization workflow, the pseudonymization 
workflow also builds on Unix standard commands to seek, generate, replace, and 
bookkeep PII from the dataset.

\pp{Workflow}
The pseudonymization workflow (Figure~\ref{fig:workflow}) is comprised of three 
independent branches, each performing series of extractions and mapping 
operations followed by substitution operations. The branching approach gives 
the workflow an ability to run each set of pseudonymization operations in 
parallel. The left branch tackles the substitution of IP addresses. The first 
stage extracts all the IP addresses from the entire dataset. The next stage 
generates pseudo-IP addresses, and finally the last stage performs the 
substitutions of the IP addresses. The middle branch performs the 
pseudonymization of user IDs from the dataset. The first stage performs the 
extraction of user IDs from the data. This extraction is performed in the same
way as for the anonymization workflows, with the addition of an extraction 
process for recording the user IDs into a single file. The next stage generates 
a list of pseudonymized user IDs and map them to real user IDs. Finally, the 
map is fed into the subsequent stage to perform the substitutions in the entire 
dataset. The right branch performs the pseudonymization of the project IDs. A 
list of active projects is also obtained from OLCF's RATS system that acts as 
a master list for project IDs to search and substitute. The first stage 
performs search in the dataset against this master list. The second stage 
generates pseudonymized project IDs, and finally the third stage performs the substitutions in the dataset. Some project IDs may appear in mixed cases in the 
dataset, thus we have carefully ensured that our substitution process properly 
handles these edge cases as well.

\pp{Experiment Conditions}
We have also implemented the pseudonymization workflow as a Swift/T workflow 
application~\cite{Maheshwari_loginanalysis_2022}. We run the workflow on the 
ORNL's Summit and Crusher leadership class HPC systems. Crusher is an OLCF's  
moderate-security system that contains identical hardware and similar software 
as the Frontier system~\cite{crusher} (the first exascale HPC system). It is 
used as an early-access testbed for the Center for Accelerated Application 
Readiness (CAAR) and Exascale Computing Project (ECP) teams as well as OLCF 
staff and the vendor partners. The system has 2 cabinets, the first with 128 
compute nodes and the second with 64 compute nodes, for a total of 192 compute 
nodes. Each compute node is equipped with 64-core AMD EPYC 7A53 ``Optimized 
3rd Gen EPYC" CPU, four AMD MI250X, each with 2 Graphics Compute Dies (GCDs) 
for a total of 8 GCDs per node with access to 64 GiB of HBM2E, 512 GB of DDR4 
memory, and connection to a 250 PB GPFS scratch filesystem\footnote{Both Summit 
and Crusher operate over the same GPFS scratch filesystem.}.
Crusher currently occupies the first position in the Green500 list 
(June 2022). By running workflow instances on both systems, we seek to contrast
the efficiency of the systems for running this category of workflow 
applications, as well as demonstrate that our implementation takes advantage
of the features provided by the system (e.g., I/O bandwidth, high-speed 
networking, processing power, etc.). To this end, we conduct scalability 
studies to understand to which extent the workflow can scale regarding data 
volume (thus identify a potential I/O bottleneck), as well as the ability of
the workflow system to handle large-scale workflows---i.e., how the workflow
system overhead may impact the efficiency of the workflow.

\begin{figure}[!t]
    \centering
    \includegraphics[width=\linewidth]{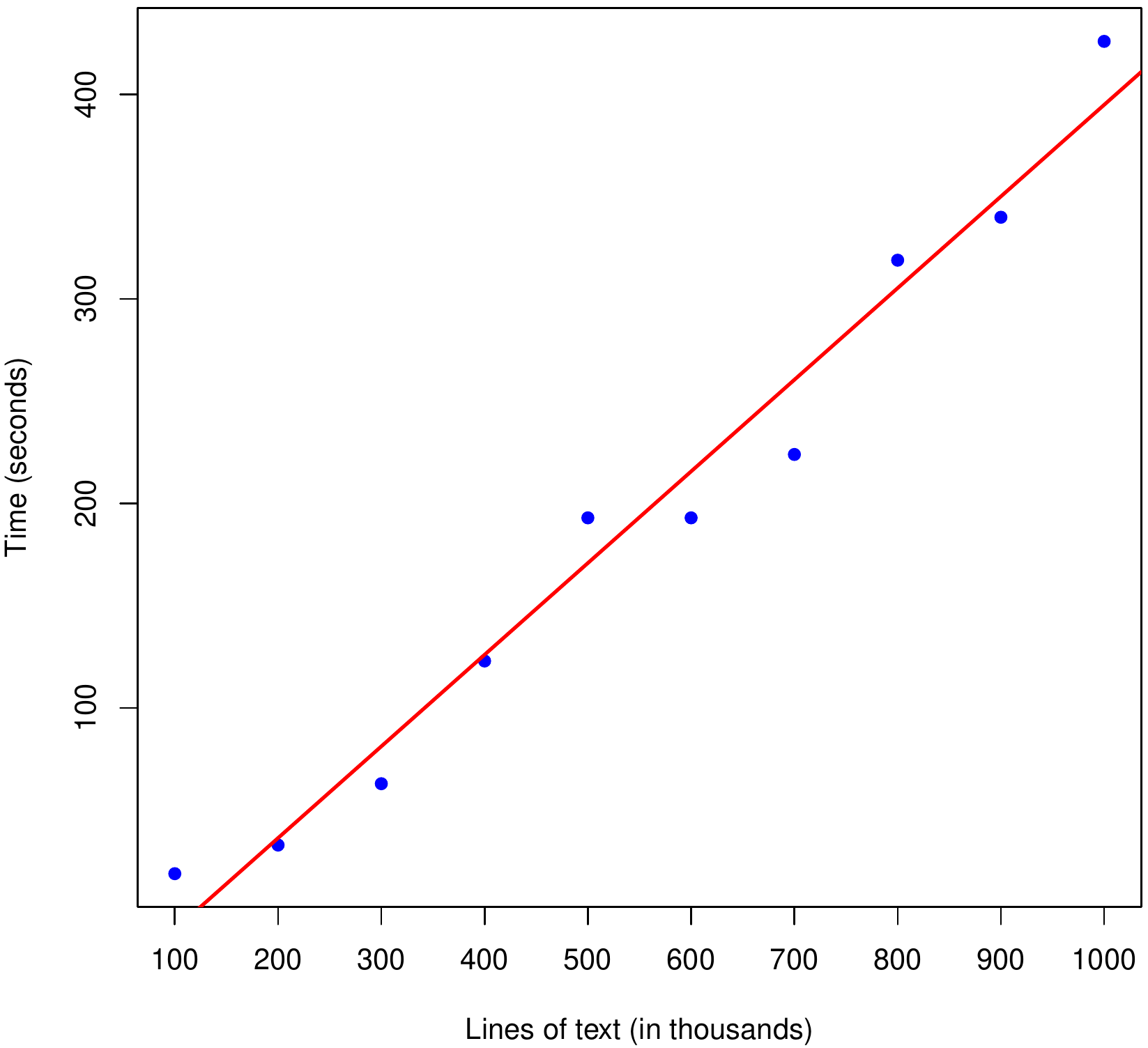}
    \caption{Pseudonymization of smaller amounts of data in serial fashion on 1 CPU as a baseline performance and estimation. (Red solid line denotes an ordinary least-squares linear regression of the execution times.)}
    \label{fig:serial}
\end{figure}

\pp{Baseline Performance}
In order to assess the need for a scalable solution, we conduct a baseline 
performance study in which we run the pseudonymization workflow serially
(using a single CPU and a single node) over small samples of data. The goal 
of this study is to estimate the time span necessary to perform the 
transformations through the entire Summit data serially and identify 
trends of the execution time when the data volume increases. The result of this
study will thus motivate the development of a parallel, scalable approach.
For this experiment, we define instances of the workflow that perform 
pseudonymization operations over determined number of lines, i.e. 100K to 1M
with increments of 100K lines on subsequent executions. Data files are choosen 
randomly based on the number of lines in each file. As the number of lines in 
the dataset files are not exactly rounded to the nearest thousand, we allow for 
a margin of 1,000 lines in each workflow configuration. 

Figure~\ref{fig:serial} shows the execution times related to increasing 
data volumes (in terms of lines of text). Execution times increase near 
linearly with the total number of lines, which demonstrates that our scripts
performs pseudonymization operations with minimum overhead. Given this result,
we extrapolate from the execution time to pseudonymize 1M lines (426 seconds)
to estimate the serial execution time of the entire Summit dataset; more than
20 hours would be required to process 175M lines of raw text serially on
a single Summit node. Although one could argue that it could still be bearable
to perform the pseudonymization process serially, we could counter this
argument as follows. First, the data collection process is a continuous effort,
thus the data volume is continuously increasing. Second, the need for
identifying potential new insights to improve HPC systems and policies fosters
the gathering of additional data, thus a potential substantial increase in the
hourly data volume. Third, HPC facilities intend to evaluate such data on a
near real-time fashion, thus having the ability to swiftly process these
datasets is of prime importance. Perhaps most importantly, OLCF policies do not
favor such long-running single-node jobs; in the default batch queue, such jobs
can only run for a maximum walltime of 2 hours, and in the killable queue, they
can run for up to 24 hours, but after the first 2 hours, they can be preempted
by higher-priority jobs.

\begin{figure}[!t]
    \centering
    \includegraphics[width=\linewidth]{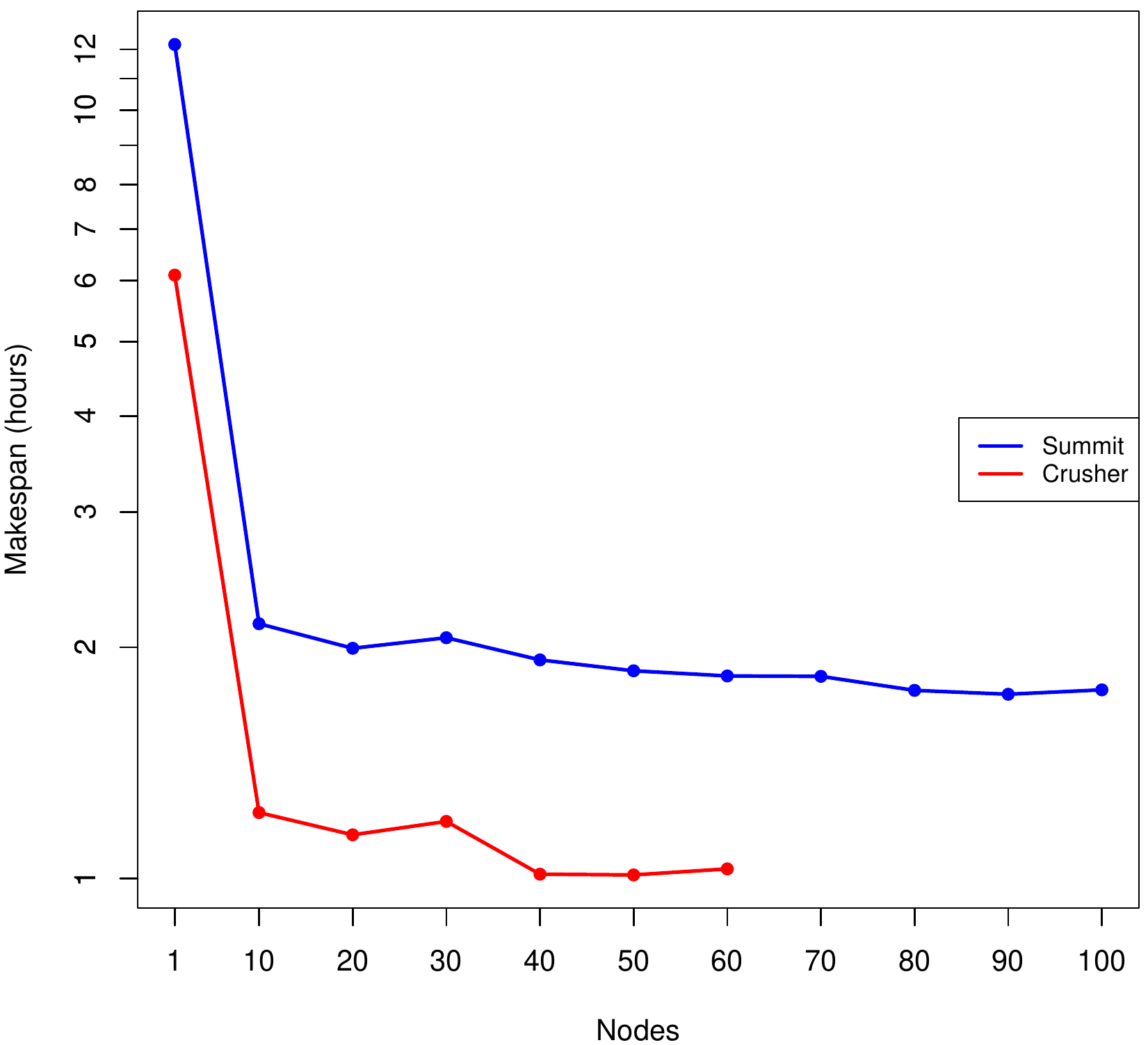} \\
    \vspace{12pt}
    \includegraphics[width=\linewidth]{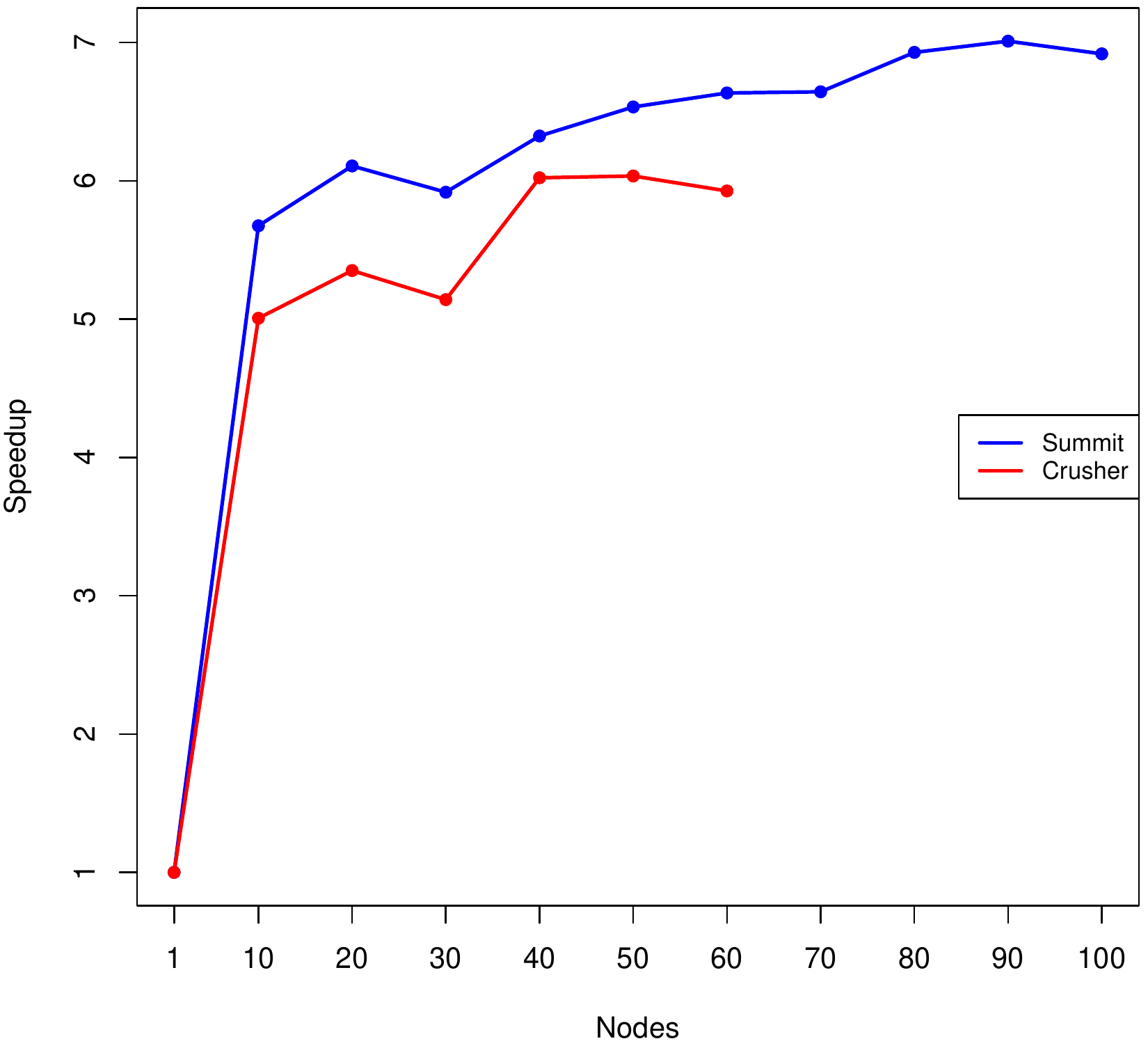}
    \caption{Pseudonymization workflow strong scaling. Runs were performed on OLCF's leadership class HPC systems: Summit (blue line) and Crusher (red line). \emph{Top:} Workflow makespan in hours; \emph{Bottom:} Workflow speedup.}
    \label{fig:strongscaling}
\end{figure}

\pp{Strong Scaling}
Figure~\ref{fig:strongscaling} shows strong scaling runs of the 
pseudonymization workflow on Summit and Crusher. We performed runs using up to
100 nodes (4,200 CPU cores) on Summit, and up to 60 nodes (3,840 CPU cores) on
Crusher. Each run processes the entire Summit data (175M lines). Not 
surprisingly, runs on Crusher yields smaller execution times (up to a factor
2, Figure~\ref{fig:strongscaling}-\emph{top}). More interesting, strong scaling 
trends on both systems are relatively similar (despite their very distinct 
architectures). We then claim that our workflow yields stable performance 
across platforms. When contrasting the execution of the workflow on a single 
Summit node (42 CPU cores) with the estimated baseline performance, we observe 
a speed up factor up to 1.6 (Figure~\ref{fig:strongscaling}-\emph{bottom}). 
While this represents a notable improvement on the execution time, the parallel 
efficiency is relatively low (0.03). 

Overall, the performance of the workflow improves with the number of nodes, 
however the parallel efficiency significantly diminishes for runs with 20+ 
nodes (regardless the HPC platform). Considering the near-linear trend from the 
baseline performance, we conjecture that the low performance is due to the 
large number of I/O operations and fine-grained data management of thousands 
of relatively small files. (This pattern is also observed in artificial 
intelligence and machine learning workflows~\cite{ferreiradasilva2021works}.)
This result indicates that effective HPC-workflows should also provide 
fine-grained data management. In future work, we plan to explore high 
performance data management frameworks for enabling in-memory processing of the
Summit data. Another limiting factor we observe is that the dataset has 3 files
that are disproportionately large than the rest of the files (up to 5 times
larger than the average) which results in a long-tail pattern on the execution
where processing those 3 files takes an additional up to 12 minutes of time. We
plan to address this in the future by incorporating a mechanism of splitting
files and processing them in parallel.

\begin{figure}[!t]
    \centering
    \includegraphics[width=\linewidth]{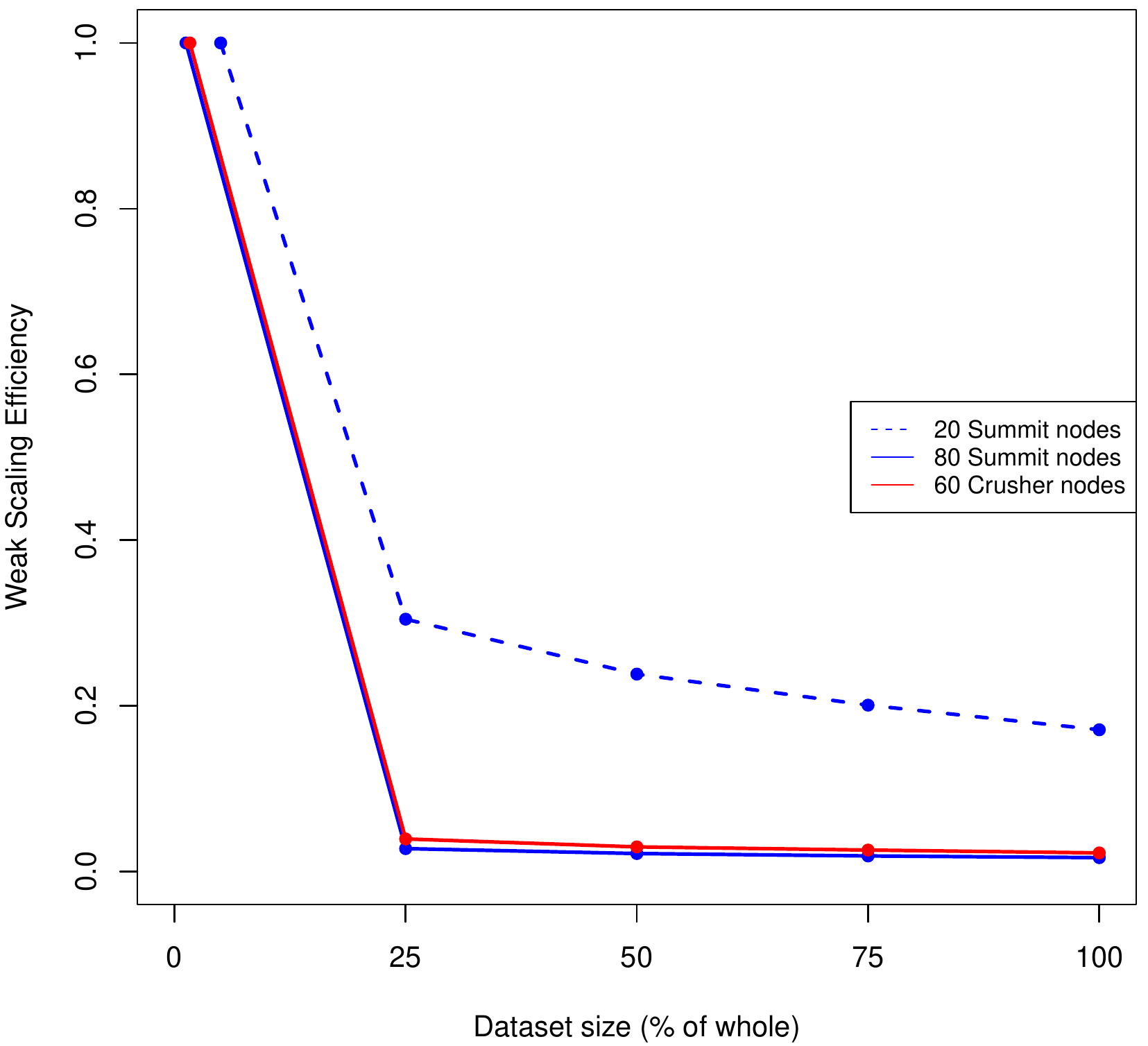}
    \caption{Pseudonymization workflow weak scaling. Runs were performed at OLCF's leadership class HPC systems: Summit (blue line) and Crusher (red line).}
    \label{fig:weakscaling}
\end{figure}

\pp{Weak Scaling}
Figure~\ref{fig:weakscaling} shows weak scaling runs of the 
pseudonymization workflow on Summit and Crusher. For these runs, we increased 
the problem size (number of lines) proportionally to the number of nodes, 
i.e. the percentage of the dataset is proportional to the percentage of nodes 
used for processing the dataset. For instance, for runs on Crusher up to 60 
nodes, we divide the dataset as follows: 1 node processes 1/60 of the dataset; 
15 nodes, 1/4 of the dataset; 30 nodes, 1/2 of the dataset; 45 nodes, 3/4 of 
the dataset, and 60 nodes, the entire dataset. Similarly, runs on Summit are
computed over 20 and 80 nodes. We arbitrarily chose 60, and 20 
and 80 as the total number of nodes (for Crusher and Summit, respectively) for
the weak scaling experiments as they yield distinct performances in the strong
scaling experiments. Our goal is to assess whether the pseudonymization 
workflow yields similar trends regardless the workflow configuration and 
problem size. Weak scaling efficiency values show similar trends, which 
corroborate with the results obtained in the baseline performance experiments.
Furthermore, the efficiency significantly drops as the more nodes are added 
and the problem size increases. This results also support our claim that the
workflow is impaired by I/O operations. Notably, the efficiency of the system
when using up to 20 nodes is nearly twice than using 60 or 80 nodes. Recall 
though that Crusher still yields considerably lower makespan at 60 nodes than 
Summit at 20 nodes (up to a factor of 2).

\medskip
Overall, we conclude that the pseudonymizing process, albeit requires 
fine-grained data management, can significantly benefit from parallel
computing techniques. Our scalability studies demonstrated that our 
implementation of the pseudonymization operations scales yield low overhead,
i.e. the limiting factor is the system capability to tackle large number 
of concurrent I/O operations over relatively small files.
A key contribution of this paper is then both the anonymization and 
pseudonymization workflows that have been publicly 
released~\cite{Maheshwari_loginanalysis_2022}. We encourage the community to 
reuse them as an efficient and parallelized solution for data 
anonymization/pseudonymization. The code may be adapted to varied needs 
including anonymization, reverse-pseudonymization, and may be repurposed as a 
scalable data processing pipeline.

\section{Challenges and Lessons Learned}
\label{sec:lessonslearned}
In this section, we discuss some of the challenges and lessons learned
surrounding technicalities of the pseudonymization process as well as the
institutional policy issues we faced during this work.

Some of these challenges were purely technical. For example, it is a challenge
to ensure consistency of user identifiers when they are occasionally truncated
by the output from Unix tools. Many tools still assume that usernames are less
than 8 characters in length, but newer OLCF users sometimes have names that
violate that assumption. Similarly, we often found user identifiers embedded
in other object names such as file and directory names; we needed to take extra
measures to ensure a mapping that preserved this information. This issue was
compounded by the fact that, for system administration reasons, special
usernames do exist which collide with common Unix tool names. Other similar
problems that we encountered included mixed-case project names as well as IP
addresses that appeared in unexpected places, which as a remedy required
looking for them across the entire dataset and not just in predefined spots.
Additionally, we needed to take care of pseudonymizing certain filepaths from
the data.

A particularly difficult question to answer with pseudonymization is, how far
is far enough? As an example, consider a hypothetical case for a user whose
name is Joseph Smith and whose username is \texttt{joesmith}. Inside his
project's shared directory, he creates a directory for his work called
\texttt{joe}, because that is what everyone on his team calls him. Our
pseudonymization workflow would replace all occurrences of his username, but
any active processes he has launched that use data from his directory might be
captured into the dataset, recording the filepath and making it possible for
analysts later to deduce that the command may have been run by a user named
``Joe''. A pseudonymization workflow might or might not know that his preferred
name is not Joseph -- that would depend on the customer management software in
use -- but even if it did, should it replace all occurrences of the string
``joe''? It turns out that ``joe'' is also the name of a text editor
\cite{joe-editor} which appears in the dataset, and this editor would have no
relation to Joseph Smith.

Thus, a major question encountered when pseudonymizing data is, how hard will
it be to relate the pseudonyms back to real identities? Because
pseudonymization preserves structure in the data, it contains more ``clues''
for parties who are interested in reverse-engineering, even though all of the
PII has been removed. This is because supplementing a pseudonymized dataset
with additional external data, such as through social engineering, can allow
for identities to be revealed; thorough anonymization so destroys the structure
of the data that deducing identities can become impossible.

These kinds of questions are of great importance at national laboratories like
ORNL, however, and our work has triggered many ongoing discussions about data
management policies for virtually all stages of the data lifecycle. One
interesting question arises from the fact that we did not use data from system
log files to construct this dataset, although we are staff members at OLCF; we
recorded this data as any ordinary user could do on a login node, using tools
that are available to everyone. This has spurred discussions about what data
our users should be allowed to see about each other, despite the words of the
ominous-sounding disclaimer that greats each new terminal session on Summit:
``Users (authorized or unauthorized) have no explicit or implicit expectation
of privacy''. Additionally, it raises questions about the need for policies on
data that can be published about the system by users. In raising these
questions about users' abilities to analyze each other, it also suggests that
there may be a set of best practices that should be publicized. One such
example already demonstrated above is not to include important or identifying
metadata in the names of directories, files, programs, scripts, and other
digital objects.

Indeed, this last point brings to mind the FAIR principles for scientific data
management and stewardship \cite{fair-principles} and discussions at ORNL about
their application to workflows \cite{reusability-first, caw2021-report,
the-f-paper}. The FAIR principles emphasize metadata practices that increase
machine-actionability, which is very helpful for constructing autonomous
scientific workflows. Recording important or identifying metadata in a filepath
is discouraged by the FAIR principles anyway.

During the course of this work, we have encountered the FAIR principles on
multiple levels. The published dataset itself is Findable through a Digital
Object Identifier (DOI), Accessible on the publicly available Constellation
system \cite{constellation}, Interoperable through representation as universal
plaintext files, and Reusable thanks to a README file that contains rich and
relevant metadata. Our workflows are Findable through a URL provided by GitHub,
Accessible using Git or a web browser, Interoperable as plaintext source files,
and Reusable as well-documented source code which has a license and full
development history. Obviously, both works can become more FAIR, and that is
something we will strive for, but thankfully, the FAIR principles are more like
guidelines than actual rules.

\section{Summary, Conclusions, and Future Work}
\label{sec:conclusion}

Understanding the needs for processing complex computational jobs is 
paramount for designing solutions that can lead to better science, and
similarly, understanding the characteristics of the user behavior on those
systems is important for improving user experiences on HPC systems. 
In this paper, we presented our experience going through the stages and 
challenges involved in managing the data lifecycle and scalable workflows at 
OLCF. In this process, we present a reusable, portable and scalable workflow 
that performs pseudonymization of a large-scale dataset. We demonstrated the 
scalability of the workflow by running weak and strong scaling experiments over 
the dataset and portability across two leadershipclass HPC architectures by 
porting it on OLCF's Summit and Crusher supercomputers. Then, we discussed 
technical and non-technical challenges and lessons learned.

In conclusion, we find that while pseudonymizing a large set of data such as 
ours is challenging, it is a worthwhile activity if done in a reusable manner 
as it will not only be useful for the community but also serve as a useful tool 
for the data as it is being produced in the context of this work and other 
activities around our institution.

We will continue to improve the workflows and finding avenues to reduce its 
complexity and execution time. One immediate approach is to bring the 
parallelism at the file level by introducing an ability to split individual 
files in such a way as to saturate large numbers of CPUs to attain better 
speeds. Another avenue that we are exploring is to port the workflow in such 
a way as to take advantage of the node local storage. We are looking into 
efficient intermediate data broadcast approaches so that the workflow stages 
running across multiple nodes may be able to access data transparently even 
if it is on node local storage.

\section*{Acknowledgments}
This research used resources of the 
Oak Ridge Leadership Computing Facility at the Oak Ridge National 
Laboratory, which is supported by the Office of Science of the U.S. 
Department of Energy under Contract No. DE-AC05-00OR22725.

\Urlmuskip=0mu plus 1mu\relax
\bibliographystyle{IEEEtran}
\bibliography{references}

\begin{thebibliography}{10}
\providecommand{\url}[1]{#1}
\csname url@samestyle\endcsname
\providecommand{\newblock}{\relax}
\providecommand{\bibinfo}[2]{#2}
\providecommand{\BIBentrySTDinterwordspacing}{\spaceskip=0pt\relax}
\providecommand{\BIBentryALTinterwordstretchfactor}{4}
\providecommand{\BIBentryALTinterwordspacing}{\spaceskip=\fontdimen2\font plus
\BIBentryALTinterwordstretchfactor\fontdimen3\font minus
  \fontdimen4\font\relax}
\providecommand{\BIBforeignlanguage}[2]{{%
\expandafter\ifx\csname l@#1\endcsname\relax
\typeout{** WARNING: IEEEtran.bst: No hyphenation pattern has been}%
\typeout{** loaded for the language `#1'. Using the pattern for}%
\typeout{** the default language instead.}%
\else
\language=\csname l@#1\endcsname
\fi
#2}}
\providecommand{\BIBdecl}{\relax}
\BIBdecl

\bibitem{mann2015allocation}
Z.~{\'A}. Mann, ``Allocation of virtual machines in cloud data centers—a
  survey of problem models and optimization algorithms,'' \emph{Acm Computing
  Surveys (CSUR)}, vol.~48, no.~1, 2015.

\bibitem{hou2022prediction}
Z.~Hou, H.~Shen, X.~Zhou, J.~Gu, Y.~Wang, and T.~Zhao, ``Prediction of job
  characteristics for intelligent resource allocation in hpc systems: a survey
  and future directions,'' \emph{Frontiers of Computer Science}, vol.~16,
  no.~5, 2022.

\bibitem{feitelson2014experience}
D.~G. Feitelson, D.~Tsafrir, and D.~Krakov, ``Experience with using the
  parallel workloads archive,'' \emph{Journal of Parallel and Distributed
  Computing}, vol.~74, no.~10, 2014.

\bibitem{javadi2013failure}
B.~Javadi, D.~Kondo, A.~Iosup, and D.~Epema, ``The failure trace archive:
  Enabling the comparison of failure measurements and models of distributed
  systems,'' \emph{Journal of Parallel and Distributed Computing}, vol.~73,
  no.~8, 2013.

\bibitem{schlagkamp2016consecutive}
S.~Schlagkamp, R.~Ferreira~da Silva, W.~Allcock, E.~Deelman, and
  U.~Schwiegelshohn, ``Consecutive job submission behavior at mira
  supercomputer,'' in \emph{25th ACM International Symposium on
  High-Performance Parallel and Distributed Computing}, 2016.

\bibitem{paul2020understanding}
A.~K. Paul, O.~Faaland, A.~Moody, E.~Gonsiorowski, K.~Mohror, and A.~R. Butt,
  ``Understanding hpc application i/o behavior using system level statistics,''
  in \emph{2020 IEEE 27th International Conference on High Performance
  Computing, Data, and Analytics (HiPC)}.\hskip 1em plus 0.5em minus
  0.4em\relax IEEE, 2020, pp. 202--211.

\bibitem{wilkinson2022iccs}
\BIBentryALTinterwordspacing
S.~R. Wilkinson, K.~Maheshwari, and R.~Ferreira~da Silva, ``Unveiling user
  behavior on {Summit} login nodes as a user,'' in \emph{Computational Science
  – ICCS 2022: 22nd International Conference, London, UK, June 21–23, 2022,
  Proceedings, Part I}.\hskip 1em plus 0.5em minus 0.4em\relax Berlin,
  Heidelberg: Springer-Verlag, 2022, p. 516–529. [Online]. Available:
  \url{https://doi.org/10.1007/978-3-031-08751-6_37}
\BIBentrySTDinterwordspacing

\bibitem{smcdc2022-website}
\BIBentryALTinterwordspacing
``Smoky mountain conference data challenge 2022,'' 2022. [Online]. Available:
  \url{https://smc-datachallenge.ornl.gov/}
\BIBentrySTDinterwordspacing

\bibitem{summitdata2022}
\BIBentryALTinterwordspacing
K.~Maheshwari, S.~Wilkinson, and R.~Ferreira~da Silva, ``Pseudonymized
  user-perspective summit login node data for 2020 and 2021,'' 5 2022.
  [Online]. Available: \url{https://www.osti.gov/biblio/1866372}
\BIBentrySTDinterwordspacing

\bibitem{gdpr}
A.~Vazao, L.~Santos, A.~Oliveira, and C.~Rabadao, ``A gdpr compliant siem
  solution,'' in \emph{European Conference on Cyber Warfare and
  Security}.\hskip 1em plus 0.5em minus 0.4em\relax Academic Conferences
  International Limited, 2021, pp. 440--XIV.

\bibitem{covidpseudo}
\BIBentryALTinterwordspacing
E.~Williamson, A.~J. Walker, K.~Bhaskaran, S.~Bacon, C.~Bates, C.~E. Morton,
  H.~J. Curtis, A.~Mehrkar, D.~Evans, P.~Inglesby, J.~Cockburn, H.~I. McDonald,
  B.~MacKenna, L.~Tomlinson, I.~J. Douglas, C.~T. Rentsch, R.~Mathur, A.~Wong,
  R.~Grieve, D.~Harrison, H.~Forbes, A.~Schultze, R.~Croker, J.~Parry,
  F.~Hester, S.~Harper, R.~Perera, S.~Evans, L.~Smeeth, and B.~Goldacre,
  ``Opensafely: factors associated with covid-19-related hospital death in the
  linked electronic health records of 17 million adult nhs patients,''
  \emph{medRxiv}, 2020. [Online]. Available:
  \url{https://www.medrxiv.org/content/early/2020/05/07/2020.05.06.20092999}
\BIBentrySTDinterwordspacing

\bibitem{Kohlmayer2019}
\BIBentryALTinterwordspacing
F.~Kohlmayer, R.~Lautenschl{\"a}ger, and F.~Prasser, ``Pseudonymization for
  research data collection: is the juice worth the squeeze?'' \emph{BMC Medical
  Informatics and Decision Making}, vol.~19, no.~1, p. 178, Sep 2019. [Online].
  Available: \url{https://doi.org/10.1186/s12911-019-0905-x}
\BIBentrySTDinterwordspacing

\bibitem{mitdata}
\BIBentryALTinterwordspacing
S.~Samsi, M.~L. Weiss, D.~Bestor, B.~Li, M.~Jones, A.~Reuther, D.~Edelman,
  W.~Arcand, C.~Byun, J.~Holodnack, M.~Hubbell, J.~Kepner, A.~Klein,
  J.~McDonald, A.~Michaleas, P.~Michaleas, L.~Milechin, J.~Mullen, C.~Yee,
  B.~Price, A.~Prout, A.~Rosa, A.~Vanterpool, L.~McEvoy, A.~Cheng, D.~Tiwari,
  and V.~Gadepally, ``The mit supercloud dataset,'' 2021. [Online]. Available:
  \url{https://arxiv.org/abs/2108.02037}
\BIBentrySTDinterwordspacing

\bibitem{bluewaters}
\BIBentryALTinterwordspacing
2022. [Online]. Available: \url{https://bluewaters.ncsa.illinois.edu/data-sets}
\BIBentrySTDinterwordspacing

\bibitem{clusterdata:Wilkes2020}
J.~Wilkes, ``Yet more {Google} compute cluster trace data,'' Google research
  blog, Mountain View, CA, USA, Apr. 2020, posted at
  \url{https://ai.googleblog.com/2020/04/yet-more-google-compute-cluster-trace.html}.

\bibitem{clusterdata:Wilkes2020a}
------, ``{Google} cluster-usage traces v3,'' Google Inc., Mountain View, CA,
  USA, Technical Report, Apr. 2020, posted at
  \url{https://github.com/google/cluster-data/blob/master/ClusterData2019.md}.

\bibitem{advpseu}
C.~Lauradoux, K.~Limniotis, M.~Hansen \emph{et~al.}, ``Data pseudonymisation:
  advanced techniques \& use cases (2021),'' 2021.

\bibitem{stalla2016anonymous}
S.~Stalla-Bourdillon and A.~Knight, ``Anonymous data v. personal data-false
  debate: an eu perspective on anonymization, pseudonymization and personal
  data,'' \emph{Wis. Int'l LJ}, vol.~34, p. 284, 2016.

\bibitem{wozniak2013swift}
J.~M. Wozniak, T.~G. Armstrong, M.~Wilde, D.~S. Katz, E.~Lusk, and I.~T.
  Foster, ``{Swift/T}: {L}arge-scale application composition via
  distributed-memory dataflow processing,'' in \emph{2013 13th IEEE/ACM
  International Symposium on Cluster, Cloud, and Grid Computing}.\hskip 1em
  plus 0.5em minus 0.4em\relax IEEE, 2013, pp. 95--102.

\bibitem{vazhkudai2018design}
S.~S. Vazhkudai, B.~R. de~Supinski, A.~S. Bland, A.~Geist, J.~Sexton, J.~Kahle,
  C.~J. Zimmer, S.~Atchley, S.~Oral, D.~E. Maxwell, V.~G.~V. Larrea,
  A.~Bertsch, R.~Goldstone, W.~Joubert, C.~Chambreau, D.~Appelhans,
  R.~Blackmore, B.~Casses, G.~Chochia, G.~Davison, M.~A. Ezell, T.~Gooding,
  E.~Gonsiorowski, L.~Grinberg, B.~Hanson, B.~Hartner, I.~Karlin, M.~L.
  Leininger, D.~Leverman, C.~Marroquin, A.~Moody, M.~Ohmacht, R.~Pankajakshan,
  F.~Pizzano, J.~H. Rogers, B.~Rosenburg, D.~Schmidt, M.~Shankar, F.~Wang,
  P.~Watson, B.~Walkup, L.~D. Weems, and J.~Yin, ``The design, deployment, and
  evaluation of the coral pre-exascale systems,'' in \emph{SC18: International
  Conference for High Performance Computing, Networking, Storage and Analysis},
  2018, pp. 661--672.

\bibitem{Maheshwari_loginanalysis_2022}
\BIBentryALTinterwordspacing
K.~Maheshwari, ``{Summit Login Nodes Usage Data Analysis},'' 2022. [Online].
  Available: \url{https://github.com/ketancmaheshwari/loginanalysis}
\BIBentrySTDinterwordspacing

\bibitem{crusher}
\BIBentryALTinterwordspacing
``Crusher,'' 2022. [Online]. Available:
  \url{https://docs.olcf.ornl.gov/systems/crusher_quick_start_guide.html}
\BIBentrySTDinterwordspacing

\bibitem{ferreiradasilva2021works}
R.~Ferreira~da Silva, H.~Casanova, K.~Chard, I.~Altintas, R.~M. Badia,
  B.~Balis, T.~a. Coleman, F.~Coppens, F.~Di~Natale, B.~Enders, T.~Fahringer,
  R.~Filgueira, G.~Fursin, D.~Garijo, C.~Goble, D.~Howell, S.~Jha, D.~S. Katz,
  D.~Laney, U.~Leser, M.~Malawski, K.~Mehta, L.~Pottier, J.~Ozik, J.~L.
  Peterson, L.~Ramakrishnan, S.~Soiland-Reyes, D.~Thain, and M.~Wolf, ``A
  community roadmap for scientific workflows research and development,'' in
  \emph{2021 IEEE Workshop on Workflows in Support of Large-Scale Science
  (WORKS)}, 2021, pp. 81--90.

\bibitem{joe-editor}
\BIBentryALTinterwordspacing
``Home -- joe's own editor,'' 2022. [Online]. Available:
  \url{https://joe-editor.sourceforge.io/}
\BIBentrySTDinterwordspacing

\bibitem{fair-principles}
\BIBentryALTinterwordspacing
M.~D. Wilkinson, M.~Dumontier, I.~J. Aalbersberg, G.~Appleton, M.~Axton,
  A.~Baak, N.~Blomberg, J.-W. Boiten, L.~B. da~Silva~Santos, P.~E. Bourne,
  J.~Bouwman, A.~J. Brookes, T.~Clark, M.~Crosas, I.~Dillo, O.~Dumon,
  S.~Edmunds, C.~T. Evelo, R.~Finkers, A.~Gonzalez-Beltran, A.~J.~G. Gray,
  P.~Groth, C.~Goble, J.~S. Grethe, J.~Heringa, P.~A.~C. ’t Hoen, R.~Hooft,
  T.~Kuhn, R.~Kok, J.~Kok, S.~J. Lusher, M.~E. Martone, A.~Mons, A.~L. Packer,
  B.~Persson, P.~Rocca-Serra, M.~Roos, R.~van Schaik, S.-A. Sansone,
  E.~Schultes, T.~Sengstag, T.~Slater, G.~Strawn, M.~A. Swertz, M.~Thompson,
  J.~van~der Lei, E.~van Mulligen, J.~Velterop, A.~Waagmeester, P.~Wittenburg,
  K.~Wolstencroft, J.~Zhao, and B.~Mons, ``The {FAIR} guiding principles for
  scientific data management and stewardship,'' \emph{Scientific Data}, vol.~3,
  no.~1, p. 160018, 2016. [Online]. Available:
  \url{https://doi.org/10.1038/sdata.2016.18}
\BIBentrySTDinterwordspacing

\bibitem{reusability-first}
M.~Wolf, J.~Logan, K.~Mehta, D.~Jacobson, M.~Cashman, A.~M. Walker,
  G.~Eisenhauer, P.~Widener, and A.~Cliff, ``Reusability first: Toward {FAIR}
  workflows,'' in \emph{2021 IEEE International Conference on Cluster Computing
  (CLUSTER)}, 2021, pp. 444--455.

\bibitem{caw2021-report}
\BIBentryALTinterwordspacing
S.~Wilkinson, K.~Knight, O.~Kuchar, K.~Mehta, M.~A. Shankar, and M.~Wolf,
  ``Official report on the 2021 {Computational and Autonomous Workflows}
  workshop ({CAW} 2021),'' United States, Tech. Rep., 2022. [Online].
  Available: \url{https://www.osti.gov/biblio/1862119}
\BIBentrySTDinterwordspacing

\bibitem{the-f-paper}
\BIBentryALTinterwordspacing
S.~R. Wilkinson, G.~Eisenhauer, A.~J. Kapadia, K.~Knight, J.~Logan, P.~Widener,
  and M.~Wolf, ``F*** workflows: when parts of {FAIR} are missing,'' 2022.
  [Online]. Available: \url{https://arxiv.org/abs/2209.09022}
\BIBentrySTDinterwordspacing

\bibitem{constellation}
S.~S. Vazhkudai, J.~Harney, R.~Gunasekaran, D.~Stansberry, S.-H. Lim,
  T.~Barron, A.~Nash, and A.~Ramanathan, ``Constellation: A science graph
  network for scalable data and knowledge discovery in extreme-scale scientific
  collaborations,'' in \emph{2016 IEEE International Conference on Big Data
  (Big Data)}, 2016, pp. 3052--3061.

\end{thebibliography}

\end{document}